\catcode`\@=11
\newif\if@fewtab\@fewtabtrue
{\count255=\time\divide\count255 by 60
\xdef\hourmin{\number\count255}
\multiply\count255 by-60\advance\count255 by\time
\xdef\hourmin{\hourmin:\ifnum\count255<10 0\fi\the\count255}}
\def\ps@draft{\let\@mkboth\@gobbletwo
    \def\@oddhead{}
    \def\@oddfoot
       {\hbox to 7 cm{$\scriptstyle Draft\ version:\ \draftdate$
       \hfil}\hskip -7cm\hfil\rm\thepage \hfil}
    \def\@evenhead{}\let\@evenfoot\@oddfoot}

\def\ceqno{\global\@fewtabfalse
    \ifcase\@eqcnt \def\@tempa{& & &}\or \def\@tempa{& &}
      \or \def\@tempa{&}
      \or\def\@tempa{}\fi\@tempa
{\rm(\theequation)}}

\def\aeqno#1{\global\@fewtabfalse
    \ifcase\@eqcnt \def\@tempa{& & &}\or \def\@tempa{& &}
      \or \def\@tempa{&}
      \or\def\@tempa{}\fi\@tempa
{\rm(\theequation,#1)}}

\def\label#1{\ifnum\draftcontrol=1
 \global\def\draftnote{$\scriptstyle #1$}\fi
 \@bsphack\if@filesw {\let\thepage\relax
   \def\protect{\noexpand\noexpand\noexpand}%
\xdef\@gtempa{\write\@auxout{\string
      \newlabel{#1}{{\@currentlabel}{\thepage}}}}}\@gtempa
   \if@nobreak \ifvmode\nobreak\fi\fi\fi
  \@esphack}

\def\alabel#1#2{\label{#1}\global\@fewtabfalse
    \ifcase\@eqcnt \def\@tempa{& & &}\or \def\@tempa{& &}
      \or \def\@tempa{&}
      \or\def\@tempa{}\fi\@tempa
{\hbox to 3cm{\phantom{\rm(\theequation,#2)}
\draftnote \hfil}\hskip -3cm {\rm(\theequation,#2)}}}

\def\clabel#1{\label{#1}\global\@fewtabfalse
    \ifcase\@eqcnt \def\@tempa{& & &}\or \def\@tempa{& &}
      \or \def\@tempa{&}
      \or\def\@tempa{}\fi\@tempa
{\hbox to 3cm{\phantom{\rm(\theequation)}
\draftnote \hfil}\hskip -3cm{\rm(\theequation)}}}

\def\eqnarray{\def\draftnote{{}}\global\@fewtabtrue
\stepcounter{equation}\let\@currentlabel=\theequation
\global\@eqnswtrue
\global\@eqcnt\z@\tabskip\@centering\let\\=\@eqncr
$$\halign to \displaywidth\bgroup\@eqnsel\hskip\@centering\@eqcnt\z@
  $\displaystyle\tabskip\z@{##}$&\global\@eqcnt\@ne
  \hskip 1\arraycolsep \hfil${##}$\hfil
  &\global\@eqcnt\tw@ \hskip 1\arraycolsep
$\displaystyle\tabskip\z@{##}$
\hfil  \tabskip\@centering&\global\@eqcnt\thr@@\llap{##}\tabskip\z@
\cr}

\def\endeqnarray{\@@eqncr\egroup
      \global\advance\c@equation\m@ne$$\global\@ignoretrue}

\def\@eqnnum{\hbox to 3cm{\phantom{\rm(\theequation)} \draftnote
                         \hfil}\hskip -3cm {\rm(\theequation)}}

\def\@@eqncr{\let\@tempa\relax
    \ifcase\@eqcnt \def\@tempa{& & &}\or \def\@tempa{& &}
      \or \def\@tempa{&}
      \or\def\@tempa{}
\fi\@tempa
\if@eqnsw
\if@fewtab\@eqnnum\fi
\stepcounter{equation}\fi\global
\@eqnswtrue\global\@eqcnt\z@\global\@fewtabtrue\cr}

\def\draftcite#1{\ifnum\draftcontrol=1#1\else{}\fi}

\def\@lbibitem[#1]#2{\item{}\hskip -3cm \hbox to 2cm
{\hfil$\scriptstyle\draftcite{#2}$}\hskip
1cm[\@biblabel{#1}]\if@filesw
     {\def\protect##1{\string ##1\space}\immediate
      \write\@auxout{\string\bibcite{#2}{#1}}}\fi\ignorespaces}

\def\@bibitem#1{\item\hskip -3cm \hbox to 2cm
{\hfil $\scriptstyle\draftcite{#1}$}\hskip 1cm
\if@filesw \immediate\write\@auxout
       {\string\bibcite{#1}{\the\value{\@listctr}}}\fi\ignorespaces}

\catcode `\@=11
\@addtoreset{equation}{section}
\def\theequation{\arabic{section}.\arabic{equation}}
\catcode `\@=12
\def\nsection#1{\section{#1}\setcounter{equation}{0}}

\def\ga{\gamma}         

\def\al{\alpha}
\def\ep{\epsilon}
\def\la{\lambda}        
\def\de{\delta}         
\def\om{\omega}

       \def\CB{{\cal B}}       
\def\CD{{\cal D}}       \def\CE{{\cal E}}       
       \def\CH{{\cal H}}       
              \def\CL{{\cal L}}
\def\CM{{\cal M}}              \def\CO{{\cal O}}

\def\qq{ \begin{eqnarray} }
\def\qqq{ \end{eqnarray} }
\def\non{ \nonumber }
\newcommand{\no}{\noindent}
\newcommand{\vs}{\vspace}

\newcommand{\p}{\partial}

\newcommand{\hf}{{_1\over^2}}
\newcommand{\ha}{{1\over 2}}

\def\draftdate{\number\month/\number\day/\number\year\ \ \ \hourmin }

\global\def\draftcontrol{0}
\catcode`\@=12

\documentclass[12pt]{article}

\usepackage{mathbbol}
\usepackage{amssymb}

\renewcommand{\theequation}{\thesection.\arabic{equation}}
\def\theequation{{\thesection.\arabic{equation}}}

\begin{document}
\begin{center}

{\Large{\bf{Approach to equilibrium for the phonon Boltzmann equation}}}

\vs{ 0.5cm}

{\large{Jean Bricmont}}\footnote{Partially supported  by the Belgian IAP program P6/02.}

UCL, FYMA, chemin du Cyclotron 2,\\ 
B-1348  Louvain-la-Neuve, Belgium\\

\vs{0.2cm}

{\large{Antti Kupiainen}}\footnote{Partially supported  by the 
Academy of Finland.}

Department of Mathematics,
Helsinki University,\\
P.O. Box 4, 00014 Helsinki, Finland\\

\end{center}
\vs{ 0.2cm}

\begin{abstract}

We study the asymptotics of solutions of the Boltzmann
equation describing the kinetic limit 
of a lattice of classical interacting anharmonic oscillators. We
prove that, if the initial condition is a small perturbation
of an equilibrium state, and vanishes at infinity, the
dynamics tends diffusively to equilibrium. The solution is the sum
of
a local equilibrium state, associated to conserved quantities
that diffuse to zero, and fast variables that are slaved to the slow ones.
This slaving implies the Fourier law, which  relates the induced
currents to the gradients of the conserved quantities.

\end{abstract}

\date{ }

\vskip 0.3cm
\begin{center}
\end{center}
\vskip 1.3 cm

\nsection{Introduction}

If a piece of solid is heated locally and left to cool, the 
initial temperature distribution will diffusively relax to a
constant temperature. The process is accompanied by
a heat flow that is proportional to the local temperature
gradient, according to the Fourier law.  A mathematical
understanding of this phenomenon starting from a
microscopic model of matter is a considerable challenge 
\cite{BLR}.

A simple  classical mechanical system modeling heat transport in solids is given by a system
of coupled oscillators
 organized on a $d$-dimensional  cubic lattice $\mathbb{Z}^d$. The oscillators are indexed by lattice
points $x\in\mathbb{Z}^d$,  and
carry momenta and coordinates $(p_x,q_x)$. In the simplest model
$p_x$ and $q_x$ take real values and their dynamics is generated
 by a Hamiltonian which is a perturbation of a harmonic system:
\qq
H (q,p) = {1\over 2} \sum_{x} p^2_{x} + {1\over 2}\sum_{xy} q_{x} q_{y} \omega^2 (x-y) +
{\la \over 4} \sum_{x} q^4_{x}.
\label{2(2)}
\qqq
In (\ref{2(2)}), 
the oscillators are coupled by harmonic forces generated by
the coupling matrix $\omega^2$ which is taken short range. The parameter
$\la$ describing the strength of the anharmonicity is assumed to be small.
The classical dynamics
 is given by Hamilton's equations
\qq
 \dot q_x=p_x ,\;\;  \dot p_{x} = -{\p H \over \p q_{x}} .
\label{2(1)}
\qqq

The dynamics (\ref{2(1)})  preserves the Gibbs measures in
the phase space, formally given by
\qq
{Z}^{-1} e^{-\beta H (q,p)} dq dp
\label{21}
\qqq
and describing an equilibrium state with constant
inverse temperature $\beta = 1/T$. One expects initial states
that agree with (\ref{21})
"at infinity" to be attracted by (\ref{21}) under the flow (\ref{2(1)}).

While such a result is well beyond current techniques,
one approach is to study this problem in an appropriate
small coupling limit, the kinetic limit. In that limit, 
one takes $\la^2=R\epsilon$, one rescales space and
time by $\epsilon$ and, in the limit as $\epsilon\to 0$, one formally
arrives at an evolution equation for the covariance
of a Gaussian measure of the form \cite{Spohn}:
\qq
 \dot{W} (x,k,t) + {1\over 2\pi} \vec \nabla \om (k) \cdot 
 \vec \nabla_{x} W(x,k,t) = RC(W)(x,k,t),
\label{c1}
\qqq
 where $x \in \mathbb{R}^d $, $k \in \mathbb{T}^d=\mathbb{R}^d/2\pi\mathbb{Z}^d$.
 The relation of $W$ to the microscopic model
 is
 \qq
 \int e^{-iky}W(x,k,t)dk=\lim_{\epsilon\to 0}(\sum_z \om(y-z)\langle q_{x/\epsilon+z}q_{x/\epsilon-z}\rangle_\epsilon+i\langle q_{x/\epsilon+y}p_{x/\epsilon-y}\rangle_\epsilon),
\label{22}
\qqq
 where $\langle-\rangle_\epsilon$  is the state at time $t/\epsilon$.
$W$ is real and
positive. Since the function $W (Rx,k,Rt) $ satisfies (\ref{c1})
with $R=1$ we may, with no loss of generality, make that assumption.

Equation (\ref{c1}) is a Boltzmann type equation where the collision term is
given by
\qq
C(W) (x,k,t) &=& {9\pi\over 4}  \int_{ \mathbb{T}^{3d}} dk_{1}dk_{2}dk_{3} (\om
\om_{1}\om_{2}\om_{3})^{-1}
 \de (\om + \om_{1}-\om_{2}-\om_{3})\non\\ &&
  \de (k+k_{1}-k_{2}-k_{3})[W_{1}W_{2}W_{3}
- W (W_{1}W_{2} + W_{1}W_{3} - W_{2}W_{3})]
\label{c2}
\qqq
 with $\om = \om (k)$, $W=W(x,k,t)$, $\om_{i} = \om (k_{i})$, $W_{i} = W (x,
k_{i},t)$, $i=1,2,3$.
The sum $k+k_{1}-k_{2}-k_{3}$ is $\textrm{mod}\  2\pi\mathbb{Z}$. We
normalize $\int_{\mathbb{T}^d} dk=1$.

Analogously to the collision term of the standard Boltzmann
equation for gases, 
the collision term (\ref{c2}) describes a scattering process.
The scattered "particles" are phonons, i.e. vibrational modes,
and the scattering process in (\ref{c2}) involves
four phonons. The phonons carry  momenta, $k, k_1, k_2, k_3$
and energies  $\om(k), \om(k_1), \om(k_2), \om(k_3)$
and the scattering process conserves the total energy and momentum.

We may pose the problem stated in the beginning, i.e.
the question of relaxation to equilibrium, for the equation
 (\ref{c1}). Thus, starting from some initial $W$ that coincides
 "at infinity" with an equilibrium state we would
 like to prove that $W(t)$ tends diffusively to
 equilibrium and the process is accompanied by
 heat fluxes governed by Fourier's law.
 In this paper, we prove such a theorem (Theorem 1 in Section 3), for an initial
 condition that is close to equilibrium. The equation
 (\ref{c1}) becomes a coupled system for
 "fast" and "slow" variables (see (\ref{Teq}), (\ref{veq})  below). The slow variables
 correspond to the temperature and the chemical potential of
 another conserved
 quantity, defined in (\ref{f5}). The slow variables flow under a nonlinear
 diffusion equation whereas the fast ones, that include
 the heat currents, are "slaved" to the slow ones by
 the Fourier law, which is discussed after stating Theorem 1 in Section 3.

 There is a rich literature on the derivation of hydrodynamics
 from the Boltzmann equation for gases. In particular in
 \cite{BU}, it is proven that, in an appropriate limit, 
  the  Boltzmann equation gives rise to the Navier-Stokes equations (see also \cite{GSR, EP, Vi} for more results and background on the link between Boltzmann and Navier-Stokes).
 In Theorem 2 (Section 3), we consider the diffusive scaling limit, i.e. the limit where one scales the spatial variable by $\ep$, and the time variable by $\ep^2$,  and we show that a suitably rescaled solution of  the  Boltzmann  equation 
 (\ref{c1}) solves, in the limit $\ep \to 0$, a
  nonlinear heat equation.

Compared with the  Boltzmann equation for gases,
eq. (\ref{c1}) has some nice features. In particular, 
its long time dynamics is described by hydrodynamic
equations that are considerably simpler than the
Navier-Stokes-Fourier system for the former (in the Euler scaling, where space and time are both scaled by $\ep$, 
eq. (\ref{c1}) reduces to the Fourier law).
A mathematical theory of (\ref{c1}) however appears to be lacking.
This paper should be considered as a step in that direction.
For an attempt to  prove the Fourier law  for phonons, 
 beyond the Boltzmann approximation, see \cite{BK, BK2}.

\nsection{The collision term}

The collision term (\ref{c2}) involves integration over the subset
determined by the constraints in the delta functions. Some assumptions
on $\om$ are needed for regularity of the resulting measure.
The standard nearest neighbor coupling between the oscillators
corresponds to $\om=\om^0$ with
\qq
\om^0(k)^2\equiv 2\sum_{j=1}^d(1-\cos k_j)+r
\label{d1}
\qqq
where $r\geq 0$ is the {\it pinning} parameter. The regularity properties
stated in Proposition 2.1 should hold for this $\om$ as long as 
$r>0$. The proof of this seems, however,  quite tedious and,
therefore, we take
\qq
\om(k)=\om^0(k)^2,
\label{d2}
\qqq
with $r>0$, which is simpler to analyze. Moreover, we need to take
the spatial dimension $d\geq 2$. {\it From now on, we'll make that assumption.}

For $r=0$ and $\om(k)=\om^0(k)$ the regularity problems are
more serious and we expect to need  $d\geq 3$.
Moreover, the spectrum of the linearized collision operator
has no gap and the nature of the flow of eq. (\ref{c1})
is quite different since also the slaved modes are diffusive.
Progress in that case is so far hampered by the 
complexity of the integrals appearing in   (\ref{c2}).

The basic properties of the collision term are summarized in Proposition 2.1.
Since the $x$ variable plays no role in $C$ we consider $C$ as a
map in 
$$\CB=C^0(\mathbb{T}^d)
$$ 
equipped with the sup norm, denoted by $\|\cdot \|$. We also need the Hilbert space
 \qq
 \CH \equiv L^2 (T^d, \om^2
(k) dk),
\label{j1}
\qqq
whose norm will be denoted by $\| \cdot \|_\CH$. In this paper, we shall use $C$ or $c$
to denote constants that may vary from place to place.
\vspace*{2mm}

\no{\bf Proposition 2.1} (a) {\it $C: \CB\to \CB$ with $\|C(W)\|\leq C\|W\|^3 $, for some constant $C<\infty$.}

\vspace*{2mm}

\no (b) {\it The equation $C(W)=0$ has, for $W\geq 0$, exactly a two parameter
family of solutions in $\CB$
\qq
W_{T,A} = {T \over \om(k)+A}
\label{c3}
\qqq
where $W\geq 0$ for $A\geq -r^2$.}

\vspace*{2mm}

\no (c) {\it  For all $W\in\CB$
\qq
 \int dk C(W) (k) = 
 \int dk\om (k) C(W) (k) = 0.
\label{c4}
\qqq}

\vspace*{2mm}

\no (d) {\it Let $-L$ be the linearization of
$C(W)$ around $W_0={\om(k)^{-1}}$ i.e.
\qq
 L = -DC(W_{0}).
\label{c8}
\qqq
Then,
 $L$ is bounded in $\CB$ and bounded and positive in $\CH $. It has two zero modes
 \qq
 L \om^{-1} = L \om^{-2}=0.
\label{c9}
\qqq
and the rest of its
spectrum in $\CB$ and  $\CH$ is contained in $\Re\lambda\geq a>0$.

Moreover $L=M+K$ where $M$ is a multiplication operator by
a strictly positive continuous function and $K$  is an integral operator
 compact in  $\CH$ and $\CB$ and satisfying}
 \qq
\sup_k\int |K(k,k')|dk'<\infty.
\label{c9a}
\qqq
 
\vspace*{2mm}

\no {\bf Proof.} 
(b)
We use an argument given in \cite{Spohn}, for a similar kernel.
First, writing the $[-]$ in (\ref{c2}) as
\qq
\prod^3_{i=0} W(k_{i}) \left[W(k_{0})^{-1} - W(k_{3})^{-1} - W (k_{2})^{-1} 
+ W (k_{1})^{-1}\right],
\label{c11}
\qqq
we have:
\qq&&
 C(W)(k)=
 {9\pi \over 4} \int_{ \mathbb{T}^{3d}} \prod^3_{i=1} (dk_{i} \om^{-1}_{i} W(k_i))
\de (\om_{0}+\om_{1}-\om_{2}-\om_{3}) \non\\
&&\cdot\de (k_{0}+k_{1}-k_{2}-k_{3})  \left[W(k_0)^{-1}- W(k_3)^{-1} -W(k_2)^{-1}+W(k_1)^{-1}
\right],
\label{cc}
\qqq
which vanishes, for  $W(k_i)\geq 0$, if $f(k)=W(k)^{-1}$ is a
{\it collisional invariant}, i.e.  a function satisfying
\qq
f(k_0)+f(k_1)=f(k_2)+f(k_3),
\label{ci}
\qqq
on the set of vectors $k_i$ in the support of the delta functions in
(\ref{cc}). In \cite{spohn2} it is proven that all $L^1$-solutions of (\ref{ci}) are of the form $a\om(k)+b$. Hence, the functions in (\ref{c3}) satisfy $C(W)=0$.

To show that these functions are the only solutions of $C(W)=0$, write, for $g$ bounded,
\qq&&
\int dk g(k) C(W)(k)=
 {9\pi \over 4} \int_{ \mathbb{T}^{4d}} \prod^3_{i=0} (dk_{i} \om^{-1}_{i} W(k_i))
\de (\om_{0}+\om_{1}-\om_{2}-\om_{3}) \non\\
&&\cdot\de (k_{0}+k_{1}-k_{2}-k_{3}) g(k_0) \left[W(k_0)^{-1}- W(k_3)^{-1} -W(k_2)^{-1}+W(k_1)^{-1}
\right],
\label{c5}
\qqq
and use the symmetry $0 \leftrightarrow 1$ of  the rest of the integrand, to replace $g(k_0)$ in front of $[-]$ 
in (\ref{c5}) by $\hf (g(k_0)+g(k_1)$, and, using the antisymmetry
of  the rest of the integrand under the exchange $0 \leftrightarrow 3$, $1\leftrightarrow
2$,  we may replace $\hf (g(k_0)+g(k_1)$ by $-\hf (g(k_3)+g(k_2))$. This implies that
(\ref{c5}) is proportional to
\qq
&&\int_{ \mathbb{T}^{4d}} \prod^3_{i=0} (dk_{i} \om^{-1}_{i} W(k_i))
\de (\om_{0}+\om_{1}-\om_{2}-\om_{3}) \de (k_{0}+k_{1}-k_{2}-k_{3})\non\\
&&\cdot \left[g(k_0)+g(k_1)-g(k_2)-g(k_3)
\right]
 \left[W(k_0)^{-1}- W(k_3)^{-1} -W(k_2)^{-1}+W(k_1)^{-1}
\right].
\label{c91}
\qqq
Taking $g(k)=W(k)^{-1}$, we obtain
\qq
&&\int_{ \mathbb{T}^{4d}} \prod^3_{i=0} (dk_{i} \om^{-1}_{i} W(k_i))
\de (\om_{0}+\om_{1}-\om_{2}-\om_{3}) \de (k_{0}+k_{1}-k_{2}-k_{3})\non\\
&&\cdot \left[W(k_0)^{-1}- W(k_3)^{-1} -W(k_2)^{-1}+W(k_1)^{-1}
\right]^2,
\label{c50}
\qqq
which vanishes, for $W(k_i)\geq 0$, only if $W(k)^{-1}$ is a
collisional invariant.

\no(c)
Since $\int dk g(k) C(W)(k)$ is proportional to (\ref{c91}), 
 it vanishes, for any $W$, whenever $g$ is a collisional invariant, in particular for $g(k)=1$ and $g(k)=\om(k)$.

\no (d) Differentiating $C(W_{T,A})=0$ with respect to $T$ and $A$ at $T=1$ and
$A=0$, we obtain
 \qq
 L \om^{-1} = L \om^{-2}=0.
\label{c90}
\qqq
Explicitely, we have (note the $-$ sign in (\ref{c8}))
\qq
 Lf &=& {9\pi \over 4\om^{2}_{0}} \int_{ \mathbb{T}^{3d}} \prod^3_{i=1} dk_{i} 
\om^{-2}_{i} \de (\om_{0}+\om_{1}-\om_{2}-\om_{3}) \de (k_{0}+k_{1}-k_{2}-k_{3})\non\\
&&\left(\om(k_{0})^2 f(k_{0}) + \om(k_{1})^2 
 f(k_{1}) - \om^2 (k_{2}) f(k_{2})\right. \left. - \om^2 (k_{3}) f(k_{3})\right).
\label{c10}
\qqq
To obtain (\ref{c10}) from (\ref{c2}), write the $[-]$ in (\ref{c2}) as in  (\ref{c11})
and expand 
\qq
W(k_{i})^{-1} = \left(\om(k_{i})^{-1} + f(k_{i})\right)^{-1} = \om
(k_{i}) - \om (k_{i})^{2}  f(k_{i}) + ...
\label{c12}
\qqq
If we take the first term in  (\ref{c12})  in the 
$[-]$, we get $0$ because of the delta function in (\ref{c2}).
So, the only linear terms in $f$ correspond to taking the second term in (\ref{c12})
inside the $[-]$, and replacing $\prod^3_{i=0} W(k_{i})$ by
$\prod^3_{i=0} \om_i^{-1}$ which implies
(\ref{c10}). Using the symmetries that led to (\ref{c50}), we see that
$(f, Lf)$ is proportional to
\qq
&&\int_{ \mathbb{T}^{4d}} \prod^3_{i=0} dk_{i} 
\om^{-2}_{i} \de (\om_{0}+\om_{1}-\om_{2}-\om_{3}) \de (k_{0}+k_{1}-k_{2}-k_{3})
\non \\
&&\left(\om(k_{0})^2 f(k_{0}) + \om(k_{1})^2 
 f(k_{1}) - \om^2 (k_{2}) f(k_{2})\right. \left. - \om^2 (k_{3}) f(k_{3})\right)^2 \geq 0.
\qqq
Thus, the zero modes of $L$ are
of the form $f=\om^{-2}g$, where $g$ are
collisional invariants, i.e.  of the form $a\om(k)+b$  (see part (b)) and therefore (\ref{c9}) give the only
zero modes in $\CH$ and in $\CB$. 

From (\ref{c10}), we get the decomposition   $L=M+K$ with
\qq
 M(k)={9\pi  \over 4\om^{2}_{0}} \int_{ \mathbb{T}^{3d}} \prod^3_{i=1} dk_{i} 
\om^{-2}_{i} \de (\om_{0} + \om_{1}- \om_{2}  - \om_{3}) \de (k_{0} + k_{1} - k_{2}
-k_{3}) \om^2 (k_{0}),
\label{c13}
\qqq
where $k=k_0$, and $K$ is  an integral operator
\[
 (Kf) (k) = \int_{ T^{d}} K (k,k') f(k') \om(k')^2dk',
\]
given by the last three terms in
(\ref{c10}), i.e. 
\qq
 K (k,k') = -{9\pi  \over 4}
(\om(k)\om(k'))^{-2} (I^{(1)}(k,k')+ I^{(2)}(k,k'))
\label{K1}
\qqq
where
\qq
&& I^{(1)} (k,k') =
 \non \\
&&  2 \int_{ \mathbb{T}^{d}}dk_1 (\om(k_1) \om(k_1+k-k'))^{-2}
\de (\om(k_1) -\om(k_1+k-k')+\om(k)-\om(k')),
\label{I1}
\qqq
and
\qq
&& I^{(2)}(k,k') =
\non \\
&& - \int_{ \mathbb{T}^{d}}dk_1 (\om(k_1) \om(k_1+k-k'))^{-2}
\de (-\om(k_1) -\om(k_1+k-k')+\om(k)+\om(k')).
\label{I11}
\qqq
The singularities of $I^{(1)}$ and $I^{(2)}$ are similar, and  
we will study  $I^{(1)}$ first; let
$$r=\hf(k-k')
$$
 and change variables in (\ref{I1})
 to $k_1=q-r- ({\pi\over 2},\dots,{\pi\over 2})$ and write 
 ${\pi\over 2}$ for $({\pi\over 2},\dots,{\pi\over 2})$. Recalling  (\ref{d1}, \ref{d2}),
we get
\qq
 I^{(1)} (k,k') =  \int_{ \mathbb{T}^{d}}dq \
\de (\Omega(q,r,k'))\ g(q,r)
\label{I2}
\qqq
with $g=(\om(q-r-\hf\pi)) \om(q+r-\hf\pi)))^{-2}$, and
\qq
 \Omega(q,r,k')=\sum_{j=1}^d(\sin(q_j+r_j)-\sin(q_j-r_j))+\om(k'+2r)-\om(k').
\label{Omega}
\qqq
The summand equals $2\cos q_j\sin r_j$. 
We change variables to $\cos q_j=s_j(1-y_j)$ with $s_j=1$ in the region $q_j\in [-{\pi\over 2},
{\pi\over 2}]$,  and
  $s_j=-1$ in the region $|q_j|>{\pi\over 2}$. 
In both cases, $y_j\in [0,1]$. Our integral
becomes a sum of integrals of the form
\qq
 I_s (k,k') = \int_{[0,1]^d} \de(2\sum_j y_j s_j\sin r_j -m(k',r))G(y,r)\prod_j
y_j^{-\hf}h(y_j)dy_j,
\label{I3}
\qqq
with  $h(y)=(2-y)^{-\hf}$,  $G$ smooth, and 
\qq
m(k',r)=\om(k'+2r)- \om(k')+2\sum_j s_j\sin r_j.
\label{m1}
\qqq
Let $\sin r_j\neq 0$ for all $j$ i.e. $k_j-k'_j\neq 0$ modulo $2\pi$ and $m(k',r)\neq 0$. Given $k$, the set of $k'$ such that
these conditions hold is of full measure.
Then,
\qq
 I_s (k,k') = \prod_j|\sin r_j|^{-\hf}J_s(k,k'),
\label{I30}
\qqq
where
\qq
 J_s (k,k') = 
 \int_0^{|\sin r_j|}\prod_j(
 {dy_j\over \sqrt y_j}h({y_j\over |\sin r_j|}))
  \de(2\sum_j y_j s_j-m(k',r))G(y_r,r),
\label{I4}
\qqq
with
 $y_{rj}=y_j/|\sin r_j|$ and $s_j=\pm 1$.

 Consider e.g. $d=2$. Integrating the delta function over the variable $y_2$ shows that
 the singularities of $J$ are the same as those of the function
 \qq
j(r,k',m)= \int_0^{|\sin r_1|}
 {\chi (sy+m)\in [0,|\sin r_2|])dy \over \sqrt y\sqrt{s y+m}},
 \label{I5}
\qqq
evaluated at $m=s'm(k',r)$ where $s,s'=\pm 1$. 
 $j$ is  bounded by
\qq
|j (r,k',m)| \leq C(1+|\log |m||).
\label{I6}
\qqq
Since $m(k',r)$ is real analytic,
 $\prod_j|\sin r_j|^{-\hf} \log m$ is square integrable
and thus
$I\in L^2(T^d\times T^d)$
and, so, $K$, given by  (\ref{K1}), is compact in $\CH$. 
From (\ref{c10}), we see that $L$ is self adjoint in $\CH$.

To show compactness in $\CB$, we will use H\"older continuity
properties of $I_s (k,k')$. We want to bound
$$
\sup_{k, \tilde k}\int dk' \frac{|I_s (k,k')-I_s (\tilde k,k')|}{|k-\tilde k|^\al},
$$
 for some $\al>0$.
We will analyze the dependence in $k$ of each factor in (\ref{I30}), (\ref{I4}) separately.
For $u,v\in (0,1]$ we have, for any $n$,
\qq
|u^{-\hf}-v^{-\hf}|=|u^{1\over 2n}-v^{1\over 2n}||\sum_{k=1}^{n}
u^{-{k\over 2n}}v^{-{n+1-k\over 2n}}|\leq C_n
|u-v|^{{1\over 2n}}
 \max\{u^{-{n+1\over 2n}},v^{-{n+1\over 2n}}\}
 \label{j0}
\qqq
since $u^{1\over 2n}\in C^{1\over 2n}([0,1])$.
We can use this, with $n>1$, to control the $k$ dependence of
 the factor $\prod_j |\sin r_j|^{-\hf}$ in (\ref{I30}).
For $J_s$, let us, for
 simplicity, again study $j$. Let $\tilde m=m(k', \tilde r)$, with $\tilde r= \ha (\tilde k-k')$.
Applying (\ref{j0}) to (\ref{I5}) we get, with $\alpha= {1\over 2n}$,
\qq
&&|\int_0^{|\sin r_1|}
 \chi (sy+m)\in [0,|\sin r_2|]) ({1\over \sqrt y\sqrt{s y+m}}-
{1 \over \sqrt y\sqrt{s y+\tilde m}}) dy |\non\\
&&\leq C_\al |m-\tilde m|^\alpha(1+\max\{|m|^{-\al},|\tilde m|^{-\al}\})\non\\
&&\leq C_\al |k-\tilde k|^\alpha(1+\max\{|m|^{-\al},|\tilde m|^{-\al}\}),
\label{dj}
\qqq
where the last inequality holds because $m$ is analytic in $k$.
Next, let $|\sin \tilde r_1|=|\sin r_1|+\de$, assuming $\de>0$. Then, using the fact that
$$\int_{|\sin r_1|}^{|\sin r_1|+\de}
 {\chi (sy+m)\in [0,|\sin r_2|])dy \over \sqrt y\sqrt{s y+m}},\leq C \min (1+ |\log |m| |, \log (1+\frac{\de }{|\sin r_1|})),
 $$
 we get the following crude estimate:
\qq
|j(r_1, r_2,k',m)-j(\tilde r_1, r_2,k',m)|&\leq& C
\de^\ha + \chi(|\sin r_1|\leq \de^\ha (1+ |\log |m| |) ).
\label{dj1}
\qqq
A similar estimate holds for the dependence with respect to $r_2$, and for the one with respect
to $m$ in $ \chi (sy+m)\in [0,|\sin r_2|]) $.
Since $m$ is real analytic,  $\prod_j|\sin r_j|^{-\hf} |m(k', \ha (k-k'))|^{-\al}$ is integrable
in $k'$ uniformly in $k$ for  $\al>0$ small enough. Moreover, the integral
of $\prod_j|\sin r_j|^{-\hf} |m(k', \ha (k-k'))|^{-\al}\chi(|\sin r_1|\leq \de^\ha )$ over $k_1'$ is of order $\de^\al$, for $\al$ small. 

We get, by
combining all these estimates, that for some $\al>0$ and all  $f \in \CB$, 
\qq
\|\int I_s(k,k')f(k')dk'\|_\al\leq C\|f\|
\label{I12}
\qqq
where $\|-\|_\al$ is the $C^\al$ norm. We may repeat this analysis for $I^{(2)}$, writing:
\qq
 &&I^{(2)}(k,k'+\pi) =
 \non \\
 && - \int_{ \mathbb{T}^{d}}dk_1 (\om(k_1-\pi) \om(k_1+k-k'))^{-2}
\de (\om(k_1) -\om(k_1+k-k')+\om(k)-\om(k')),
\label{I13}
\qqq
where we changed variables $k_1 \to k_1+(\pi, \dots, \pi)$ in the integrand and used
formula (\ref{d1}), (\ref{d2})  for $\om (k)$ and $\cos (k_j+\pi)=-\cos k_j$. Since the factor $(\om(k_1-\pi) \om(k_1+k-k'))^{-2}$ is smooth,
we can do exactly the same analysis as above and obtain the bound (\ref{I12}) for $I^{(2)}$.
Then, 
the same bound holds
for the kernel $K$, given by (\ref{K1}).This shows that $K$
is a continuous map from $\CB$ into the space $C^\al$
of H\"older continuous functions. Since $C^\al$
is compactly embedded in $\CB$ (by  Ascoli's  theorem), this proves that $K$ is compact. 
Since, using (\ref{K1}), (\ref{I1}), (\ref{I11}), (\ref{I13}),
$$M(k)=\int dk' K(k,k')\om(k')^2,$$ 
we also get $M\in\CB$
(actually $M$ is H\"older continuous) and that $L$ is bounded (since $K$ is compact).

The case  $d>2$ is similar.

\no (a) We proceed as in (d) with the delta functions and get
\qq
|C(W)(k)| \leq C\|W\|^3\int dk'
 \int_{[0,1]^d} \de(2\sum_j y_j s_j\sin r_j -m(k',r))\prod_j
y_j^{-\hf}h(y_j)dy_j
\label{I7}
\qqq
which, using (\ref{I30}), (\ref{I5}), proves the claim.
\hfill $\Box$

\vspace*{4mm}

The identities  (\ref{c4}) yield two conservation laws. To show this,
define, for $\al=1,2$,
\qq
j_\al(x,t)=-{1\over 2\pi}(\om^{-\al},\nabla\om \ W),
\label{c161}
\qqq
where the scalar product is in $\CH$, see (\ref{j1});
$j_1$ is the thermal current and $j_2$ can be called the 
phonon number current.
 Similarly, set
 \qq
T_\al(x,t)=(\om^{-\al},W).
\label{c162}
\qqq
$T_1$ is the temperature and $T_2$ is related to the
phonon chemical potential.
Equations (\ref{c1}) and  (\ref{c4}) give then the conservation laws
  \qq
\dot{T}_\al=\nabla\cdot j_\al.
\label{c163}
\qqq
$T_\al$ are the slow modes that will diffuse. The currents
$j_\al$ will be related to their gradients via the Fourier law, see (\ref{c182}) below.

\nsection{Results}

Let $\hat W(p,k,t)$ denote the Fourier transform of $W$
in the $x$ variable. We shall look for solutions of (\ref{c1}) of the form:
\qq
 \hat W(p,k,t) = W_0(k)\delta(p)+ w(p,k,t).
\label{c16}
\qqq
where, we recall, $W_0=\om^{-1}$.
The equation  (\ref{c1}) becomes then,
 \qq
  \dot{w}=-Dw+n(w),
\label{c16a}
\qqq
where the linear operator is given by
 \qq
 D=L+{_i\over^{2\pi}}p\cdot \nabla\om(k)
\label{c16b}
\qqq
and the nonlinear term
is $n(w)=C(W_0+w)+Lw=C(W_0+w)-DC(W_0)w $. Written as an integral equation, 
(\ref{c16a}) becomes
 \qq
 w(t)=e^{-tD}w(0)+\int_0^t ds e^{-(t-s)D}n(w(s)).
\label{c16c}
\qqq
We need to decompose this in terms of the slow 
and the fast variables. 
Let $P$ be the orthogonal projection in
the Hilbert space  $\CH$ on 
$E=
\textrm{span} \ \{ \om^{-1}, \om^{-2}\}$
and let $Q$ be the one on
the complement of $E$. The identities
(\ref{c4}) can then be written as:
\qq
Pn=0,
\label{c15}
\qqq
or $n=Qn$, since, by differentiation, (\ref{c4}) implies the same identities
 with  $C(W)$ replaced by $DC(W_0)w$.
Let 
$$Pw=T, \ \ Qw=v.$$
Then, $w(t)=T(t)+v(t)$, and eq. (\ref{c16a}) becomes
 \qq
  T(t)&=&Pe^{-tD}w(0)+\int_0^t ds Pe^{-(t-s)D}Qn(w(s)) \label{Teq}\\
 v(t)&=&Qe^{-tD}w(0)+\int_0^t ds Qe^{-(t-s)D}Qn(w(s))
\label{veq}
\qqq
Define
\qq
e(p,t):=(1+(t+1)p^2)^{-n}
 \label{g1}
 \qqq
with $n>d/2$ and let $\CE_t$ be the space of continuous functions $f(p,k)$
equipped with the norm
\qq
\|f\|_t \equiv \sup_{k,p}\ e(p,t)^{-1} |f(p,k)|=\sup_{p}\ e(p,t)^{-1}  \|f(p,\cdot)\|
\label{ca18}
\qqq
and let $\CE$ be the space of functions $w(t)\in\CE_t$, with
\qq
\|w\|_\CE \equiv \sup_t \|w(t)\|_t.
\label{cb18}
\qqq
Let $\kappa:E\to E$ be the linear operator 
\qq
\kappa=(2\pi )^{-2}P\partial_1\om L^{-1} \partial_1 \om P,
 \label{c1812}
 \qqq
 where $\partial_1$ denotes the derivative with respect to the first argument. 
 $\kappa$ is strictly
positive by Proposition 2.1.d, since  the set $\partial_1\om E$
forms a two dimensional subspace in $E^\perp$ (by symmetry: $\partial_1\om$ is odd in $k$, while
the functions in $E$ are even). Let
\qq
T_0(p,t)=e^{-tp^2\kappa}T(p, 0)\chi(|p|\leq 1),
\label{f1}
\qqq
and
\qq
v_0(p,t)=  {-i\over 2\pi} L^{-1} p \cdot\nabla \om  T_{0} (p,  t) 
\label{f2}
\qqq
(these lie in $E$ and $E^\perp$ respectively).
Then, we have our main result:

\vspace*{2mm}


\no{\bf Theorem 1.}  {\it There exists $\de >0$ such that, for $\|w(0)\|_0 \leq \de$,
the equation}  (\ref{c16a}) {\it  has a unique solution in $\CE$, $w(t)=T(t)+v(t)$,
satisfying, for $t\geq 1$,
\qq
\|T(t)-T_0(t)\|_t \leq C(t)t^{-\hf}  \|w(0)\|_0,\label{f3}\\
\|v(t)-v_0(t)\|_t\leq C(t)t^{-1} \|w(0)\|_0,
\label{f4}
\qqq
where $C(t)=C\log (1+t)$ for $d=2$ and  is constant for $d>2$.}

\vspace*{2mm}

\no{\bf Remark.}  The norm (\ref{ca18}) and (\ref{f3}) imply that, in $x$ space, $T(x,t)$
is given by the usual diffusive Gaussian term, which, for bounded $x$, decays like
$t^{-{d\over2}}$, plus a correction bounded, uniformly in $x$, by $\CO(t^{-{d+1\over 2}})$.
Eq. (\ref{f4}) says that the fast mode
$v$ is slaved to the slow one $T$, i.e. it decays like an explicit term, the Fourier transform of $v_0(p,t)$,
 plus a correction uniformly bounded by 
$\CO(C(t) t^{-{d+2\over2}})$.

Eq. (\ref{f4}) also implies the Fourier
law for the leading terms of the solution. Write $T$
in the basis
\qq
T(p,k, t) =\sum_{\beta=1,2} \om^{-\beta} (k) \tilde T_\beta (p,t),
\label{f5}
\qqq
(since the basis is not orthogonal, $\tilde T_\beta$ does not coincide with the Fourier transform of
$T_\beta$ in  (\ref{c162})).
Then, in $x$-space, the currents
 (\ref{c161}) (where, by symmetry, only the $v$ part of $W$ contributes) become, up to terms  $\CO(C(t)t^{-{d+2\over2}})$,
\qq
j_{\al}=\sum_\beta\kappa_{\al\beta} \nabla \tilde T_\beta,
\label{c182}
\qqq
with the positive conductivity matrix
\qq
\kappa_{\al\beta} =(\om^{-\al},\kappa\om^{-\beta}),
\label{c183}
\qqq
since, by symmetry, $(\om^{-\al},\partial_i \om L^{-1}\partial_j \om \om^{-\beta})$
equals zero for $i\neq j$, and, for  $i=j$, $i$ can be chosen equal to $1$.

\vspace*{2mm}

Finally, we can also derive a nonlinear heat equation as the hydrodynamic scaling limit
of the  Boltzmann equation  (\ref{c1}). 
We scale
\qq
\tilde W (x,k,t) = W (\ep x, k, \ep^2 t),
\non
\qqq
where $\tilde W (x,k,t)$ satisfies (\ref{c1})). Thus, we obtain, for $W$, the equation:
\qq
\dot W (x, k, t) + (2\pi \ep)^{-1} \nabla \om(k)\cdot \nabla W (x,k,t) = \ep^{-2} C(W) (x, k, t).
\label{51}
\qqq
We shall solve it with initial data $W |_{t=0}= \om (k)^{-1}+w (x, k, 0)$ and
\qq
w (x,k, 0) = T (x,k, 0) + \ep v (x,k, 0),
\label{52}
\qqq
with $T (x, \cdot, 0) \in E$, $ v_{0} (x, \cdot, 0) \in E^\perp$, $\forall x$.
$\tilde W (x,k,0) = W (\ep x,k, 0)$ has spatial variations at scale
$\ep^{-1}$. 

We have then 

\vskip 0.4cm

\no{\bf Theorem 2.}   {\it  There exists $\de >0$ such that, for all $\ep \leq 1$
and $\|T (0)\|_0$, $\|v (0)\|_0 \leq \delta$, the equation (\ref{51}) has a unique
solution $W^\ep = T^\ep + \ep v^{\ep} \in \CE$.
Moreover, $T^\ep (t) \to T(t)$ and $v^{\ep} (t) \to v(t)$ in $\CE_{t}$, for all
$t>0$, as $\ep \to 0$, where $T$ and $v$ are the unique solutions of
\qq
DC(\om^{-1} + T) v = (2\pi)^{-1} \nabla \om \cdot \nabla T \label{v}\\
\dot T = -(2\pi)^{-1} P \nabla \om \cdot \nabla v
\label{T}
\qqq}

\vspace*{2mm}

\no{\bf Remark.} 
Since we have $DC(\om^{-1} + T)$ instead of $DC(\om^{-1})=-L$  in (\ref{T}), we get a nonlinear heat equation for
$T$:
\qq
\dot T = \nabla\cdot({\cal K}(T)\nabla T)\label{T'}
\qqq
with 
\qq
{\cal K}(T)= -(2\pi)^{-2}P\partial_1\om DC(\om^{-1} + T)^{-1}\partial_1\om P,
\qqq
(remembering that $L$ is positive, we see that  $DC$ is negative and that ${\cal K}(T)$ is a positive matrix).
 One can show that the long-time behavior of (\ref{T'}) is similar to the one in Theorem 1.

\nsection{Proofs}

We start by proving some results on the spectrum of
the operator $D$ defined in (\ref{c16b}).
 Recall that, by Proposition 2.1 d, $\Re(\sigma (L) \backslash \{0\})\geq a$.

\vspace*{2mm}

\no{\bf Proposition 4.1} {\it There exists $p_{0}>0$, and  $b>0$, such that, both
in $\CB$ and in $\CH$, the following holds:} 

\no a) {\it If $|p| < p_{0}$
then
\[
\sigma (D) = \{\lambda_{1},\lambda_{2}\} \cup \tilde \sigma 
\]
with $ \inf\ \Re \tilde \sigma > a/2$ and
\qq
\lambda_{i} = p^2 \mu_{i} + \CO (|p|^3)
\label{l1}
\qqq
where $\mu_{i} >0$ are the eigenvalues of  the operator
$\kappa$ defined in }(\ref{c1812}). 

\no b) {\it If $|p| \geq p_{0}$ then $\inf \Re\sigma (D)  \geq b$.}

\vspace*{2mm}

\no {\bf Proof.} a) Let
\[
L(\beta) = L + \beta \rho
\]
where $\rho=(2\pi)^{-1}|p|^{-1}p\cdot \nabla\om(k)$
so that $D=L(i|p|)$. $\rho$ is a bounded operator in $\CB$ 
and in $\CH$. Hence, for $|\beta| < p_{0}$ small enough, the
spectrum of $L(\beta)$ consists of two eigenvalues
close to the origin, with the rest of the spectrum having real part larger than
$a/2$. Moreover, 
 $L(\beta)$ is self adjoint in $\CH$ for $\beta$ real and thus forms an analytic Kato family (see \cite{RS}). Therefore, there is a $p_0$
such that, for $|\beta| < p_{0}$,  the two eigenvalues  of $\CL (\beta)$, $\lambda_{1}$, $\lambda_{2}$, and the associated 
eigenfunctions $\psi_{i}$, are analytic in $|\beta| < p_{0}$. To compute their Taylor
series, use the decomposition $w=T+v$ and project the equation $(L+\beta \rho)w = \lambda w$
onto $Q\CH$ and $P\CH$. Then, using $P \rho T=0$ (since $\rho T$ is odd in $k$),
 we get a pair of equations:
\qq
&&Lv+\beta Q  \rho (T+v) = \lambda v,\non\\
&&\beta P \rho v = \lambda T,
\non
\qqq
whereby, since  $\Re(\sigma (L) \backslash \{0\})\geq a$, we get, for $\beta$ small
and $\la =\CO(\beta)$,
$$
v=-\beta (L-\la+\beta Q\rho)^{-1} Q \rho T =-\beta L^{-1} Q \rho T + \CO ( \beta^2)T,
$$
where $\CO ( \beta^2)$ denotes a bound on the operator norm in $\CB$.
Since
$Q \rho T= \rho T$ and $T=PT$,  we get 
$$-\beta^2 P \rho L^{-1} \rho PT+ \CO
(\beta^3)T=\lambda  T.$$ The $2\times 2$ matrix $P\rho L^{-1} \rho P$ is strictly
positive, by Proposition 2.1 (d).  The result follows by letting $\beta =i |p|$, and by observing that, since the functions in $E$ are invariant under reflections and permutations of the coordinates of $k$,  the matrix $\kappa(p)=P  \rho L^{-1} \rho P$ equals $\kappa(e_1)$ i.e. the one given in (\ref{c1812}).

\vspace*{2mm}

\no b)  Write  $D = \CM +i|p|\rho+K$ as a sum of a multiplication and a compact
operator. Since $\Re\sigma (\CM + i|p| \rho) \geq \inf \CM>0$ we need only to show that all
eigenvalues $\lambda$ satisfy Re$\lambda  \geq b$, if $|p|
\geq p_{0}$. We decompose $\CH = \CH_{e} \oplus \CH_{o}$ into even and odd subspaces.
Let $\lambda$ be an eigenvalue. Then, since $\rho$ is odd, and the operator $L$ is
 parity-preserving,
\qq
Lw_{e} + i |p| \rho w_{o} = \lambda w_{e}\non\\
Lw_{o} + i|p| \rho w_{e} = \lambda w_{o}
\label{e1}
\qqq
Write $\lambda=x+iy$ and suppose that $x<{a\over 2}$ (where $a=\inf \sigma (L
 \mid_{\CH_{o}}))$. Then $(L-\lambda) \mid_{\CH_{o}}$ is invertible, and
(\ref{e1})
becomes
\qq
(L+p^2 \rho (L-\lambda)^{-1}\rho) w_{e} = \lambda w_{e}.
\label{ee2}
\qqq
Let $\|w_{e}\|_\CH = 1$. Then, taking the scalar product of  (\ref{ee2}) in $\CH$ with $w_e$, and 
writing separately the real and imaginary parts, one gets:
\qq
x = (w_{e}, (L+p^2 \rho (L-x)((L-x)^2+y^2)^{-1} \rho) w_{e})\non\\
\geq {a\over 2} p^2 (w_{e},  \rho ((L-x)^2 + y^2)^{-1} \rho w_{e}),
\label{e3}
\qqq
since $L\geq 0$ and, as operators in $\CH_{o}$, $(L-x)((L-x)^2+y^2)^{-1}\geq {a\over 2}((L-x)^2+y^2)^{-1}$.
For the imaginary part, we get:
\qq
y=yp^2 (w_{e}, \rho ((L-x)^2+y^2)^{-1} \rho w_{e}).
\label{e4}
\qqq
If $y\neq 0$ (\ref{e3}) and (\ref{e4}) imply $x \geq a/2$, against
our assumption.
Thus $y=0$, i.e. all eigenvalues $\lambda$ of $\CD$ with Re$\lambda<{a\over
2}$ are real, and, by (\ref{e3}), positive.

\vspace*{2mm}
Let
\[
l(r,\lambda) = \inf_{\|w\|=1} (w, (L+r^2 \rho (L-\lambda)^{-1} \rho)w).
\]
Since, for $\la \leq a/2 $, $(L-\lambda)^{-1}\geq c >0$, we get
$$(w, (L+r^2 \rho (L-\lambda)^{-1} \rho)w)   \geq
 (w,(L+cr^2 \rho^2)w).$$ 
 To bound this from below, observe that both terms in
 $L+cr^2 \rho^2$ are positive,  that 
 $(w,Lw) \geq c\|Qw\|^2 $, and that
  $(Pw, \rho^2 Pw)\geq c\|Pw\|^2 $, by 
  inspection
   of the functions in $P\CH$ and of $\rho$.
  So, if $\|Qw\|^2 \geq c' \|w\|^2 $, we can use the first lower bound, while, if 
  $\|Qw\|^2 \leq c' \|w\|^2 $, i.e. $ \|Pw\|^2\geq (1-c') \|w\|^2 $, we can use the second one, if $c'<<c$, to get
 $ L+cr^2\rho^2 > c''r^2$ for $ r<p_0$, and $p_0$ small enough.
 So, $l(p,\lambda) \geq c''p^2$, for $|p|\leq p_0$, and $\lambda \leq a/2$. 
Besides,  $l(r,\lambda) \leq l(r',\lambda) $ if $r<r'$. Since by (\ref{ee2}) $\lambda
\geq l(|p|,\lambda)$ we conclude $\lambda \geq c''p^2_{0}$, for $|p| \geq p_0$. Thus, we can take $b=\min ({a\over 2}, c''p^2_{0})$ \hfill $\Box$

\vspace*{2mm}

We are now ready to state the estimates for the heat kernels in equations
  (\ref{Teq}) and (\ref{veq}). Let 
  $$R(z)=(z-D)^{-1}$$
   be the resolvent
  of the operator $D$ and $\ga$ be a circle of radius $a/4$ around the origin. 
Then 
  \qq
\tilde P= \oint_{\ga} R(z)\  {_{dz}\over^{ 2\pi i}},
\label{tildep}
\qqq
is a projection onto the 2-dimensional
eigenspace of $D$  introduced in Proposition 4.1 a.      Let $\tilde Q=1-\tilde P$. 
We have:

\vspace*{1mm}

\no{\bf Proposition 4.2} {\it For all $t\geq 0$ we have 
\qq
\| e^{-tD}\|                                \leq C (e^{-ctp^2}+e^{-ct})\label{421}
\qqq
where $\|\cdot\|$ is the operator norm in  $\CB$. Moreover,
there exists $p_0>0$ so that  for $|p| \leq p_{0}$
\qq
\|Pe^{-tD}Q\| + \|Qe^{-tD}P\| &\leq&  C|p| (e^{-ctp^2} + e^{-ct})\label{422}\\
\| Q e^{-tD} Q \|                          &\leq& C (p^2e^{-ctp^2} + e^{-ct})\label{423}\\
\|  e^{-tD} \tilde Q \|                          &\leq& C  e^{-ct}\label{424}
\qqq
}

\vspace*{2mm}

\no {\bf Proof.}  Since $D=L+{_i\over^{2\pi}}p\cdot \nabla\om(k)$, we have, by the resolvent expansion:
\qq
\tilde P  
&=&   \oint_\ga (z-L)^{-1} \  {_{dz}\over^{ 2\pi i}}  +
 \oint_\ga (z-L)^{-1}{i \over 2\pi} p \cdot \nabla \omega  
(z-L)^{-1}\  {_{dz}\over^{ 2\pi i}} + \CO (p^2)
\non\\
&=& P- {i\over 2\pi} L^{-1} p \cdot \nabla \om P - {i\over 2\pi} P p \cdot \nabla \om  L^{-1} Q
  + \CO (p^2)
  \non\\
&=& P+ A P + PBQ
  + \CO (p^2)
\label{e61}
\qqq
where,  here and below, $\CO (p^2)$ denotes a bound on the norm of operators in $\CB$, and
we define
\qq
A=-{i\over 2\pi} L^{-1} p \cdot \nabla \om,\;\;\; B=-{i\over 2\pi} p \cdot
\nabla \om L^{-1}.
\label{l2}
\qqq
Hence, for $|p|$ small,
\qq
\| P \tilde Q  \| + \| \tilde Q  P\| +\| \tilde P Q \| \ +  \|Q \tilde P \| \leq C |p|.
\label{*}
\qqq
 Now, write
\qq
e^{-tD} =  \tilde P e^{-tD} \tilde P + \tilde Q e^{-tD} \tilde Q,
 \label{j2}
\qqq
(since $\tilde P $, and $\tilde Q$ project on  invariant subspaces of $D$, we have $e^{-tD} \tilde P=\tilde P e^{-tD} \tilde P$, $e^{-tD} \tilde Q=\tilde Q e^{-tD} \tilde Q$),
and
\qq
 e^{-tD} \tilde Q=\oint_\ga e^{-tz} 
  R(z)\  {_{dz}\over^{ 2\pi i}}\label{dunf}
\qqq
where the curve $\ga$ 
goes around the part of the spectrum of $D$ that lies in the complement of a ball of radius $a/4$, centered at the origin. Since $L$ is bounded and $|p|$ small, the length of $\ga$ is $\CO(1)$, and we get  (\ref{424}).
 Then, 
the other claims follow, for $|p|$ small, from (\ref{*}),  (\ref{j2}) and Proposition 4.1 a. 

To get (\ref{421}) for  $|p|>p_0$ we note that
by Proposition 4.1, $e^{-tD}$ is given by the right hand side of (\ref{dunf}),
where the curve $\ga$ can be chosen to be  a rectangle in $\Re z>b$, with vertical
sides of length ${\rm const}\ |p|$, and horizontal ones of length $\CO(1)$ (since $L$ is bounded).
Hence, on the curve $\ga$, we have $|e^{-tz}|\leq e^{-b t}$, but the length of the contour
of integration is
not bounded as $|p|\to \infty$. To control the integral, recall that  $D=M+i|p|\rho+K$, and
let $$R_0(z)=(z-M-i|p|\rho)^{-1}.$$
Then, by the resolvent formula,
\qq
R=R_0+R_0KR_0+R_0KR_0KR.
 \label{reso}
\qqq
The first term in  (\ref{reso}), inserted  in  (\ref{dunf}), gives
the multiplication operator $e^{-t(M+i|p|\rho)}$, which satisfies the
bound b) with $c=\inf M$. For the second  term, we use $|e^{-tz}|\leq e^{-b t}$
in the integral and write
\qq
|\oint_\ga dk' dz R_0(z,k)K(k,k')R_0(z,k')| \leq \int  dk' |K(k,k')| \oint_\ga dz {1\over |z-z_1||z-z_2|}
 \label{reso1}
\qqq
where $z_1=(M+i|p|\rho) (k)$, $z_2= (M+i|p|\rho) (k')$. 
Since
$ |z-z_i|\geq c$ on $\ga$, we have
\qq
\oint_\ga dz {1\over |z-z_1||z-z_2|}\leq C,
 \label{reso2}
\qqq
uniformly in $k,k'$.
Since, moreover, $\int |K(k,k')|dk'\leq C$, uniformly in $k$ (which holds
by  (\ref{c9a})), we get the bound in b) for the second term in (\ref{reso}).
 The third term in (\ref{reso}) is similar, since $KR$ is bounded.
\hfill $\Box$

\vspace*{4mm}
\no {\bf Proof of Theorem 1.} Let us write  eq. (\ref{c16c}) as 
\qq
w(t)=w_{\ell}(t)+N(w,t)
\label{w(t)}
\qqq
where $w_{\ell}(t)$ is the solution of the linear problem:
\qq
w_{\ell}\equiv e^{-tD} w (0).
\label{w^o}
\qqq
We will solve  (\ref{w(t)}) in the space $\CE$.

The linear term is bounded, using   (\ref{421}), and recalling the definition (\ref{g1}) of the weight
function $e$, by
\qq
\|w_{\ell}\|_\CE\leq C \sup_{p,t} e(p,t)^{-1}e(p,0)(e^{-ctp^2}+e^{-ct}) \| w (0)\|_0 \leq  C\| w (0)\|_0.
\label{linear}
\qqq

Consider then the nonlinear term in eq. (\ref{c16c}). By an easy extension of
Proposition 2.1 a (see (\ref{I7})),
\qq
\|n(w(s)(p)\| \leq C(\|w\|_\CE^{2} e(s)^{*2}(p)+\|w\|_\CE^{3} e(s)^{*3}(p)).
\label{j9}
\qqq
We need the simple estimate
\qq
e(s)^{*2}(p)\leq C(1+s)^{-{_d\over^2}}e(p, s),
\label{e2}
\qqq
which follows from
\qq
e(s)^{*2}(p)\leq 2\int_{|p-q|\geq \hf |p|} dq \ e(p-q,s)e(q,s)\leq
Ce(p,s)\int dq\ e(q,s),
\label{jb}
\qqq
(the first inequality holds because  either  $|q|\geq \hf |p|$ or $|p-q|\geq \hf |p|$).
The last integral yields the power of $1+s$ since
$n>\hf d $ in (\ref{g1}). The last term in (\ref{j9}) is smaller, for $\|w\|_\CE\leq 1$.

Thus, the second term in equation (\ref{Teq}) and the second term in (\ref{veq})
  are bounded as
by
\qq
\|RN(w,t,p)\|\leq C\|w\|_\CE^{2} \int_0^tds\|Re^{-(t-s)D}Q\|(1+s)^{-{_d\over^2}}e(s,p)
\label{vT}
\qqq
where $R$ equals  $P$ in   (\ref{Teq}) and $Q$ in (\ref{veq}). Proposition 4.2
implies
\qq
\|Re^{-(t-s)D}Q\|\leq C((1+t-s)^{-\hf m}e^{-{c\over 2}(t-s)p^2}+e^{-c(t-s)}),
\label{jb1}
\qqq
with $m=1$ if $R=P$ and $m=2$ if $R=Q$ (using $|p|^me^{-{c\over 2}p^2(t-s)}\leq C(1+t-s)^{-\hf m}$, for $t-s\geq 1$). Bounding
$$
e^{-{c\over 2}p^2(t-s)}e(s,p)\leq Ce(t,p)
$$ 
and 
$$e^{-{c\over 2}(t-s)}e(s,p)\leq Ce(t,p)
$$
we may bound  the remaining $s$-integral in  (\ref{vT}), 
for $m=1, 2$, i.e. for $m\leq d$, by
$
C(t)(1+t)^{-\hf m}
$
where $C(t)$ may be taken to be  constant, except
for $d=2$, 
where $C(t)=C\log(1+t)$. Hence we end up with
\qq
\|RN(w,t)\|_t \leq C(t)(1+t)^{-\hf m}\|w\|_\CE^{2}.
\label{jb2}
\qqq
In particular, we have 
$$
\|N(w)\|_\CE\leq C \|w\|_\CE^{2},
$$
so  that $N$ in (\ref{w(t)}) maps a ball of radius $\epsilon$
in $\CE$ into itself, for $\epsilon$ small enough and is obviously
a contraction there. The existence of a solution for eq.  (\ref{w(t)})
then follows from (\ref{linear}) and
 the Banach fixed point
theorem and, since  $m=1$ if $R=P$, i.e. for 
(\ref{Teq}),  and $m=2$ if $R=Q$  i.e. for 
(\ref{veq}), we obtain the bounds 
\qq
\|T(t)-T_{\ell}(t)\|_t \leq C(t)(1+t)^{-\hf}  \|w(0)\|_0^2,\label{f3a}\\
\|v(t)-v_{\ell}(t)\|_t\leq C(t)(1+t)^{-1} \|w(0)\|_0^2,
\label{f3aa}
\qqq
where we wrote
\qq
w_{\ell}=Pw_{\ell}+Qw_{\ell}\equiv T_{\ell}+v_{\ell}.
 \label{we} 
\qqq

To conclude the proof of Theorem 1 we need to 
relate  $T_0$, $v_0$  to $T_{\ell}$, $v_{\ell}$. For this we need to write
the leading terms of $w_{\ell}(t)$ more explicitely.
Let us formulate this a little more generally, which will be useful also
in the proof of Theorem 2. Let us denote
\qq
K(t)=e^{-tp^2\kappa}P               
\label{ko}
\qqq
where $\kappa$ is given in eq, (\ref{c1812}).
We have then
 the following Lemma,
proven at the end:

\vspace*{1mm}

\no{\bf Lemma 4.3}  {\it Let $|p|\leq p_0$. Then the semigroup $e^{-tD} $
can be written
with respect to the decomposition $E\oplus E^\perp$, as
\qq
e^{-tD} =\left(\begin{array}{ccccc}
K(t)&  K(t)B\\
AK(t)& A K(t)B   +R(t) 
\end{array}\right)    
    +\left(\begin{array}{ccccc}
\CO(|p|e^{-ctp^2}+p^2e^{-ct})  & \CO(p^2e^{-ctp^2}+|p|e^{-ct})\\
 \CO(p^2e^{-ctp^2}+|p|e^{-ct})&  \CO(|p|^3e^{-ctp^2})
\end{array}\right)  
\non
\qqq
where $\CO$ is with respect to the operator norm in $\CB$,
\qq
  R(t) &=&Q\tilde Q   e^{-tD}    \tilde QQ
  \label{K11} 
\qqq
and the operators $A$ and $B$ are defined in equation} (\ref{l2}).

\vspace*{2mm}

Returning to the proof of Theorem 1 and recalling the
definition (\ref{f1}) of $T_0$ and  (\ref{w^o}),  (\ref{we}) of $T_{\ell}$, we get  from Lemma 4.3, for $|p| \leq p_{0}$,
\qq
\| T_{\ell}(t,p) - T_0(t,p)\|\leq C|p|(
e^{-ctp^2}  + e^{-ct})e(p,0) \|w(0)\|_0.
\label{j7}
\qqq
Since $v_0$ given in eq.  (\ref{f2}) equals, see (\ref{l2}),  $v_0=AK(t)T(0)$
for $|p| \leq p_{0}\leq 1$, we have, from Lemma 4.3,
\qq
\| v_{\ell}(t,p) -v_0(t,p)\|\leq C(p^2
e^{-ctp^2}  + e^{-ct})e(p,0) \|w(0)\|_0
\label{j6}
\qqq
where we used the bound, coming from (\ref{424}), 
\qq
\|R(t)\|\leq   C e^{-ct}.
\label{K1b}
\qqq
For $|p| \geq p_0$,  by (\ref{421}),
$$\|w_0(t,p)\| +\| w_{\ell}(t,p)\| \leq Ce^{-ct} e(p,0)\|w(0)\|_0.$$ 
Using now the simple estimates
$$e(p,t)^{-1}(e^{-ctp^2/2} +e^{-ct/2})\leq Ce(p,0)^{-1},$$ for $t\geq 1$, $e(p,t)^{-1}\leq Ce(p,0)^{-1}$
for $t\leq 1$,  and 
\qq
|p|^me^{-ctp^2/2}+e^{-ct/2}\leq C(m)t^{-\hf m},
\non
\qqq
for $t\geq 1$, $m\geq 0$,
we conclude that
\qq
\|T_{\ell}(t)-T_0(t)\|_t &\leq& C(1+t)^{-\hf} \|w(0)\|_0,\label{f3b}\\
\|v_{\ell}(t)-v_0(t)\|_t&\leq& C(1+t)^{-1} \|w(0)\|_0.\label{f3c}
\qqq
The estimates (\ref{f3}) and (\ref{f4}) follow now
from (\ref{f3a}), (\ref{f3aa}),  (\ref{f3b}) and  (\ref{f3c}).
\hfill $\Box$

\vspace*{4mm}
\no {\bf Proof of Theorem 2.} The proof goes along the lines of the one of Theorem 1 and we will be brief.
We expand
\qq
{1\over \ep^2} C(W) =- {1\over \ep^2} Lw+{1\over \ep} m (T,v) + n (T,v),
\non
\qqq
where $m (T,v)$ is the term quadradic in $T$ but linear in $v$, and $n (T,v)$ collects the terms
that are quadratic and cubic in $v$. Note that, since $LT=0$,
${1\over \ep^2} Lw= {1\over \ep} Lv$ and  thus
\qq
DC(\om^{-1} + T) v=-Lv + m(T,v).
\label{l3}
\qqq
We write   (\ref{51}) as
\qq
w(t) = \exp({-{t\over \ep^2} D_{\ep}}) w(0) + {1\over \ep} \int^t_{0} ds \exp({-{t-s
\over \ep^2} D_{\ep}}) (m(s)+\ep n(s)) ds,
\non
\qqq
where
\qq
D_{\ep} = L + \ep {i\over 2\pi} p \cdot \nabla \om,
\no
\qqq
i.e. it equals (\ref{c16b}) evaluated at $\ep p$.

The decomposition $w=T+\ep v$ yields, as in (\ref{Teq}), (\ref{veq}) :
\qq
&T(t) =&
Pe^{-t D_{\ep}/\ep^2} PT(0) + \ep Pe^{-t D_{\ep}/\ep^2}
Qv(0)+\non\\&&
 {1\over \ep} \int^{t}_{0} ds Pe^{-(t-s) D_{\ep}/\ep^2} Q(m(s)+\ep n (s))
\label{54}
\\ 
&v(t) &= {1\over \ep} Qe^{-t D_{\ep}/\ep^2} PT(0)+Qe^{-t D_{\ep}/\ep^2}
 Qv(0) 
 +\non\\&& {1\over \ep^2} \int^t_{0} ds Qe^{-(t-s) D_{\ep}/\ep^2}
Q(m(s)+\ep n (s))
\label{54a}
\qqq

Consider first the case $ |p|\leq p_0/\ep$. We use Lemma 4.3, 
with $t$ replaced by $t/\ep^2$ and $p$ by $\ep p$.
This leads to
\qq
T&=&T^*+  T_1
\label{58}\\
v &=& AT^*  + {1\over \ep^2} \int^t_{0} ds Q\tilde Q e^{-(t-s) D_{\ep}/\ep^2}  \tilde Q Q
m (s) + v_{1}
\label{57}
\qqq
where
\qq
T^*=e^{-t p^2\kappa}  T(0) +  \int^t_{0} ds
e^{-(t-s)p^2\kappa} P Bm (s),
\label{59a}
\qqq
and 
$$
T_1=  \ep Pe^{-t D_{\ep}/\ep^2} Qv(0)+
 \int^{t}_{0} ds Pe^{-(t-s) D_{\ep}/\ep^2} Q n (s)+\int^{t}_{0} ds 
 \CO(\ep  p^2e^{-c(t-s)p^2}+ |p|e^{-c(t-s)/\ep^2}) m (s),
 $$
 $$
 v_1= Qe^{-t D_{\ep}/\ep^2} Qv(0) +
 {1\over \ep} \int^t_{0} ds Qe^{-(t-s) D_{\ep}/\ep^2}Qn(s) + \ep \int^t_{0} ds\CO ( |p|^3 e^{-c(t-s)p^2}) m(s).
 $$
 Then, following the proof of (\ref{f3b}), (\ref{f3c}),  (\ref{f3a}),  (\ref{f3aa}), we get
\qq
\|T_1(t,p)\|\leq C\ep (1+t)^{-\hf}e(p,t)(\|T(0)\|+\|v(0)\|),
\label{60}\\
\|v_1(t,p)\|\leq C(t)\ep (1+t)^{-1}e(p,t)(\|T(0)\|+\|v(0)\|),
\label{59}
\qqq
where $C(t)=C\max (t^{-1/2}, \log (1+t))<\infty$, for $t>0$.
We need to study the second term on the RHS of eq. (\ref{57}).
Write $m(s)=m(t)+(m(s)-m(t))$ and consider the first term
\qq
{1\over \ep^2} \int^t_{0} ds Q\tilde Q e^{-(t-s) D_{\ep}/\ep^2}  \tilde Q Q
m (t)&=&{1\over \ep^2} \int^\infty_{0} ds Q \tilde Q e^{-s D_{\ep}/\ep^2}  \tilde Q Q
m (t) + \CO(   e^{-c t/\ep^2})  \non\\
&=&Q \tilde QD_{\ep}^{-1} \tilde Q Qm(t)+\CO( e^{-c t/\ep^2})
\non\\ 
 &=&L^{-1} m(t) +\CO(\ep)+\CO(  e^{-c t/\ep^2})
\label{62}
\qqq
where we used $Q \tilde QD_{\ep}^{-1} \tilde Q Q=QL^{-1}Q+\CO(\ep)$
and  where $\CO(\cdot) $ denotes a bound on the norm
$\sup_{| p|\leq p_0/\ep} e(p, t)^{-1}\|Ê\cdot \|$ of the remainders, since $\sup_{| p|\leq p_0/\ep} e(p, t)^{-1}\|Êm(t) \|$ is bounded, uniformly in $\ep$, because of the bounds on $T$, $v$, coming from 
(\ref{58})-(\ref{59}).

For the second term,  we use the equations (\ref{54}) and (\ref{54a}), to show that 
 $\|Êm(s)-m(t) \|_t \leq  C |t-s|$, for $|t-s|$ small; whence
\qq
 {1\over \ep^2} \int^t_{0} dsQ \tilde Q e^{-(t-s) D_{\ep}/\ep^2}  \tilde Q Q
(m (s) -m(t))=\CO( \ep) .
\label{61}
\qqq
Altogether we obtain
\qq
v(t)=AT^*(t)+L^{-1}m(t)+v_2
\label{63}
\qqq
and, for any $t>0$,
\qq
\sup_{ |p| \leq p_0/\ep} e(p, t)^{-1}(\|T_1 \|+\|v_2 \|)\to 0
\label{64}
\qqq
as $\ep \to
0$.
For $|p|> p_{0}/\ep$ we have, from Proposition 4.2, 
\qq
\| e^{-t D_{\ep}/\ep^2} \| \leq e^{-ct/\ep^2},
\non
\qqq
which implies that 
\qq
\sup_{ |p| \geq p_0/\ep} e(p, t)^{-1}(\|T \|+\|v \|)\to 0,
\label{640}
\qqq
as $\ep \to
0$.
Hence, in the space $\CE_t$, we have
\qq
\lim_{\ep\to 0}(T(t),v(t))=(T^*(t), AT^*(t)+L^{-1}m(t))\equiv (T^*(t), v^*(t))
\label{641}
\qqq

Let us finally check that 
$T^*$ and $v^*$ satisfy the equations  (\ref{v}) and (\ref{T}).
From (\ref{l3}) and (\ref{641}), we get
\qq
AT^*=-L^{-1}DC(\om^{-1}+T^*)v^*
\label{642}
\qqq
which is  (\ref{v}) if we recall the definition  (\ref{l2}) of $A$.

 From (\ref{59a}) and  the definition (\ref{l2}) of $B$, we see that $T^*$
satisfies
\qq
\dot T^* = -p^2 \kappa T^* - {i\over 2\pi} Pp\cdot \nabla \om
 L^{-1} m.
\label{T1}
\qqq
From  (\ref{641}) and the definition  (\ref{l2}) of $A$, we have
$$
  L^{-1} m=v^* -A T^*=v^*  +  i(2\pi)^{-1} L^{-1} p\cdot\nabla \om
T^* .
$$ 
Substituting this into  (\ref{T1}) and recalling
the definition  of $\kappa$  in (\ref{c1812}), which implies that
$ -p^2 \kappa T^* + p^2 (2\pi)^{-2} P\cdot \nabla \om L^{-1} \cdot\nabla \om
T^* =0$, we get 
 equation (\ref{T}).  \hfill $\Box$

\vs{4mm}

\no {\bf Proof of Lemma 4.3.} We need to calculate $Re^{-tD}R'$
for $R$ and  $R'$ either  $P$ or  $Q$. We use the representation  (\ref{j2}) for $e^{-tD}$
and the formula  (\ref{e61}) for $\tilde P$. We need the following expansions
\qq
P\tilde P &=& P+PBQ + \CO (p^2), \;\;\;\;  \;\;\;\;\;\;   \;\;\;\;\;\;  \;\;\;\;\;\;  \;\;\;\;\;\;\;\tilde P P = P+ AP + \CO (p^2)
\non\\
\tilde P Q &=& PBQ + \CO (p^2), \;\;\;\;  \;\;\;\;\;\;  \;\;\;\;\;\;  \;\;\;\;\;\;  \;\;\;\;\;\;  \;\;\;\;\;\;  \;\;
Q \tilde P = AP+ \CO (p^2)
\non\\
P \tilde Q &=& P(1-\tilde P) = - PBQ + \CO (p^2),  \;\;\;\;  \;\;\;\;\;\;  \;\;\;\; \tilde Q  P= - Q AP+ \CO (p^2)
\non\\
\tilde Q Q &=& (1-\tilde P) Q = Q- PBQ + \CO ( p^2),  \;\;\;\;  \;\;\;\;\;\; Q \tilde Q = Q -  AP+ \CO (p^2)
\non
\label{55}
\qqq
where we used repeatedly $PAP=0$ ($A$ maps $E$ into $E^\perp $, because $\nabla \om$ is odd in $k$).
Thus we get,
 using the identities on the left, 
\qq
&&Pe^{-tD}Q = P\tilde P e^{-tD} \tilde QQ + P \tilde Qe^{-tD} \tilde Q Q \non\\
&& =  (P+ PBQ)e^{-tD}PBQ -  PBQ e^{-tD}
(Q- PBQ)+ \CO ( p^2 e^{-ct p^2})\non\\
&&=  e^{-t p^2\kappa} PBQ + \CO ( p^2 e^{-ct p^2})+ \CO ( |p|
e^{-ct})
\label{56}
\qqq
where we use the bounds of Proposition 4.2, $B=\CO(|p|)$, and $\la_{\al} = \mu_{\al} p^2 + \CO (p^3)$, 
with $\mu_\al$ being the eigenvalues of $\kappa$, see (\ref{l1}).

Similarly, we get
\qq
Qe^{-tD} Q &=&  Ae^{-tp^2\kappa} P
BQ  + Q\tilde Q e^{-tD}\tilde Q Q
+\CO (|p|^3 e^{-ctp^2})
\non
\qqq
and
\qq
P e^{-tD}P &=& e^{-tp^2\kappa}P 
+ \CO
( |p| e^{-ctp^2}) + \CO (p^2e^{-ct}) \non\\
Qe^{-t D}P &=&A e^{-tp^2\kappa}P + \CO (p^2 e^{-ct p^2})+\CO ( |p|
e^{-c t}).
\non
\qqq
\hfill$\Box$

\vspace*{4mm}

\no {\bf {Acknowledgments.}}
We thank  Fran\c cois Huveneers, Jani Lukkarinen and Herbert Spohn
for useful discussions. J. B. thanks the Belgian Internuniversity Attraction Poles Program and A.K.  thanks the Academy of Finland
for financial support.

\end{document}